\documentclass[preprint,12pt,authoryear]{elsarticle}

\usepackage{amsthm}
\usepackage{amsmath}

\usepackage{natbib}
\usepackage{amssymb}
\usepackage{makecell}
\usepackage{graphicx} 
\usepackage{subfigure} 
\usepackage{float}
\usepackage{color}
\usepackage{booktabs}

\begin{document}

\begin{frontmatter}



\title{Coevolutionary dynamics of population and institutional rewards in public goods games}


\author[label1]{Shijia Hua}
\author[label1]{Linjie Liu\corref{cor1}}
\cortext[cor1]{Corresponding author} \ead{linjieliu1992@nwafu.edu.cn}

\affiliation[label1]{organization={College of Science, Northwest A \& F University},
            city={Yangling},
            postcode={712100}, 
            country={China}}

\begin{abstract}
In social dilemmas, individuals face a conflict between their own self-interest and the collective interest of the group. The provision of reward has been shown to be an effective means to drive cooperation in such situations. However, previous research has often made the idealized assumption that rewards are constant and do not change in response to changes in the game environment. In this paper, we introduce reward into the public goods game and develop a coevolutionary game model in which the strength of reward is adaptively adjusted according to the population state. Specifically, we assume that decreasing levels of cooperation lead to an increase in reward strength, while increasing levels of cooperation result in a decrease in reward strength. By investigating coevolutionary dynamics between population state and reward systems, we find that interior stable coexistence state can emerge in our model, where the levels of cooperation and reward strength remain at constant levels. In addition, we also reveal that the emergence of a full cooperation state only requires a minimal level of reward strength. Our study highlights the potential of adaptive feedback reward as a tool for achieving long-term stability and sustainability in social dilemmas.
\end{abstract}




\end{frontmatter}


\section{Introduction}

Achieving cooperative action among multiple agents is a challenging task, especially in scenarios where individual interests are at odds with the common good \citep{Perc_PR_17,Zhangis22,Oh2001is,liu22is,Hua_PhysicaD_2023,Quan21kbs,Xia_IEEE_23,Xia_Auto_23}. This is particularly true in social dilemmas, such as public goods provision or shared resource management, where individuals may be inclined to free-ride on others' contributions instead of making their own \citep{han2015,han2021,han2022,han2022interface,Tanimoto2015,Tanimoto2021,Szolnoki2014interface,Ito2018RSOS}. To overcome this challenge and promote cooperation, effective incentive mechanisms are required that incentivize individuals to cooperate while minimizing the negative consequences of free-riding behavior \citep{Sigmund2010,Henrich2006,wang2019cnsns,Chen2015pre,wang2019cnsns,Cimpeanu_DGA_2023}. Negative incentives, such as costly punishment or social exclusion, may be necessary to discourage free-riders and promote equitable distributions of costs and benefits among group members \citep{Szolnoki2017prx,Chen2014njp}. 
However, the use of negative incentives can also lead to unintended negative consequences that reduce overall social utility. For instance, the imposition of costly punishment or social exclusion may trigger retaliation or anti-social punishment behaviors, leading to a breakdown of cooperation among group members \citep{Herrmann2008,Rand2011,Hua_CSF_2023}. As a result, recent research has focused on developing positive incentive mechanisms, such as rewards, that encourage cooperation through mutual benefit and shared purpose, while avoiding the negative consequences of punitive measures.

The use of reward has been shown to be an effective means of promoting cooperation in social dilemmas, providing tangible benefits for those who choose to cooperate while discouraging free-riders from exploiting the efforts of others \citep{du2018epjb,Balliet2011,Szolnoki2010}. Previous theoretical research has proposed various types of rewards, including peer reward \citep{Ozono2020sr,Hauert2010jtb} and institutional reward \citep{li2021csf,Szolnoki2015prsb,Dong2019prsb,sun2021iscience,Sasaki_12_PNAS}. In a peer incentive scenario, individuals have the ability to reward others at a personal cost. Previous research has shown that peer rewards can curb free-riding behavior in human populations. However, the implement of this type of reward system is costly, which has raised concerns about the emergence of second-order free-riders. In the institutional reward scenario, individuals are not responsible for implementing reward. Instead, an institution rewards them based on their contributions to the group. Institutional rewards can overcome the problem of second-order free-riders. However, maintaining an institution is more expensive than implementing peer reward, as participants still have to pay for its upkeep even if no individual is being rewarded \citep{Dong2019prsb,sun2021iscience}.

Although previous research has revealed the role of different types of reward in the evolution of cooperation, many of these studies have made an idealized assumption that the strength of rewards remains constant regardless of changes in the game environment, which cannot accurately reflect many real-world scenarios. For example, suppose that a city government decides to offer rewards to residents who participate in a garbage cleaning activity to encourage more people to join. If only a few people are willing to participate, the government may offer higher reward amounts or other benefits to attract more people. However, as the number of participants increases, the government may find that its budget cannot afford such high rewards, and may therefore reduce the strength of reward to avoid overspending. Until now, there has been a lack of theoretical research on how the strength of reward adapts to changes in the proportion of cooperators in the population. It remains unclear whether adaptive reward can promote the emergence of cooperation. The question of how much reward strength is needed to maintain high levels of cooperation is still worth exploring. 

Here, we establish a feedback-evolving game model in which the strength of reward and population status form a two-way feedback loop. Specifically, changes in the level of cooperation will alter the strength of rewards, which in turn affects individual decision-making. We adopt a linear feedback form to describe the effect of strategies on reward strength in public goods games by following previous works \citep{Weitzpnas2016,Szolnoki2018,wang2020,chen2018,Shao2019,Tilman2020}. Through analyzing the dynamics of the coupled system, we find that the system can exhibit a stable coexistence state, in which both the frequency of cooperators and the strength of reward can be maintained at a fixed level. Furthermore, we find that a state of high-level cooperation and even full cooperation can be achieved with relatively low levels of reward strength.

\section{Related works}

The study of cooperation evolution is one of the most important research topic in artificial intelligence and social science \citep{Perc_PR_17,Zhangis22}. It is well-known that cooperation among individuals can provide benefits to the group as a whole, but it also faces the challenge of cooperation dilemma, where individual rationality conflicts with collective rationality. To overcome this dilemma, various mechanisms have been proposed and studied. One such mechanism is reward, which provides positive feedback to individuals who exhibit cooperative behavior \citep{Szolnoki2010,Dong2019prsb}. The use of rewards has been shown to be effective at promoting cooperation in many contexts, ranging from traditional social dilemmas like the prisoner's dilemma to modern online communities and marketplaces.

In recent years, a variety of reward forms have been proposed to promote cooperation and enhance social welfare in different contexts. Peer reward, as mentioned earlier, is one such incentive form that provides rewards to others at a cost to themselves. Previous behavioral experiments and theoretical studies have revealed the pivotal role played by peer reward in the evolution of cooperation \citep{Gneezy12prsb,Sigmund01pnas,Hilbe10prsb}. However, one of the biggest obstacles to the evolutionary stability of the peer reward is the presence of second-order free-riders, i.e., individuals who abstain from dispensing such incentives. In addition to being costly, the success of peer reward is also challenged because it may be co-opted to promote antisocial behavior. Some individuals abuse the reward opportunity by engaging in antisocial reward, which increases the welfare of free-riders and, in turn, harms cooperators.

Institutional reward can overcome the problem of second-order free-riders and avoid antisocial behaviors \citep{sun2021iscience,Sasaki_12_PNAS}. When this mechanism is effective, individuals no longer bear the cost of rewarding cooperators themselves, but rather cooperators are rewarded through an institutional. However, this approach is more wasteful than peer-based incentives because individuals must pay the cost to maintain the institution even if no one receives a reward. Theoretical studies based on evolutionary game theory suggest that the impact of institutional reward on cooperation can be understood from the perspective of the size of incentives \citep{Sasaki_12_PNAS,Chen_15_interface}. When the size of incentives is very small, institutional rewards do not promote cooperation, and defection is the most advantageous strategy. If the incentive value is at an intermediate level, cooperators and defectors can coexist stably in the population. For larger incentive values, cooperators will dominate the entire population. Moreover, recent theoretical research has unveiled that institutional reward can be employed to foster the emergence of fairness within the context of ultimatum games \citep{Fangis2020,Cimpeanu_CSF_2023}.

Although institutional reward plays an important role in the evolution of cooperation, previous research has typically assumed that the strength of incentives is constant and does not vary with changes in the game environment. For unlimited incentive budgets, providing high rewards naturally promotes the emergence of high-level cooperation. However, in actual social systems, incentive budgets are typically limited, making it particularly important to design an adaptive and changing reward intensity. Here, we propose an adaptive reward mechanism in which the intensity of reward is linearly related to the number of cooperators in the game group. Specifically, the more cooperators there are, the lower the intensity of reward, and vice versa. 

\section{Model and Methods}

\subsection{Public goods games with institutional reward}

We examine a well-mixed population that is infinitely large, where groups of $N$ players are randomly formed to engage in a public goods game. In this game, each player decides whether to contribute $c$ to a commonpool or keep it for themselves. Individuals who choose to contribute are referred to as cooperators, while individuals who do not contribute are referred to as defectors. The total contributions are multiplied by a factor $r$ and then evenly distributed among all group members, regardless of whether they contributed or not. Therefore, defectors always receive a higher payoff from the public goods game than cooperators, leading to the social dilemma.

To address the aforementioned social dilemma, we introduce institutional reward into public goods games. Building on previous work \citep{Chen_15_interface}, we consider a total budget of rewards $N\delta$ in each game group, where $\delta$ denotes the average per capita reward. The entire incentive budget is allocated to reward cooperators equally within the game group. As for cooperators, their payoffs are increased by $\frac{aN\delta}{N_{C}}$, where $a$ represents the intensity of reward, and $N_{C}$ represents the number of cooperators among the remaining $N-1$ players, excluding the player itself. Consequently, the payoffs for a cooperator and defector in the public goods game can be expressed as follows:
\begin{eqnarray}
\pi_{C}&=&\frac{rc(N_{C}+1)}{N}-c-\delta+\frac{aN\delta}{N_{C}+1}, \\
\pi_{D}&=&\frac{rcN_{C}}{N}-\delta.
\end{eqnarray}

\subsection{Replicator dynamics}

We use replicator equations to describe how the proportion of cooperators evolves over time \citep{Schuster_83,Hofbauer_98}. Specifically, the change in the frequency of cooperators is proportional to the difference between the average payoff of cooperators and the average payoff of the population. The proportion of cooperators will increase if their average payoff is higher than that of defectors, and vice versa. Let $x$ and $1-x$ be the proportions of cooperators and defectors, respectively. Accordingly, we have
\begin{eqnarray}\label{equation 1}
\dot{x} &= x(1-x)(f_{C}-f_{D}),
\end{eqnarray}
where $f_{C}$ and $f_{D}$ denote the average payoffs of cooperators and defectors, respectively. In an infinite well-mixed population, the average payoffs of cooperators and defectors can be written as
\begin{eqnarray}\label{equation 2}
   f_{C}&=&\sum_{N_{C}=0}^{N-1}\binom{N-1}{N_{C}}x^{N_{C}}(1-x)^{N-N_{C}-1}\pi_{C},\\
   f_{D}&=&\sum_{N_{C}=0}^{N-1}\binom{N-1}{N_{C}}x^{N_{C}}(1-x)^{N-N_{C}-1}\pi_{D},
\end{eqnarray}
where $\pi_{C}$ and $\pi_{D}$ represent the payoffs of cooperators and defectors in the game, respectively, as defined by Equations (1) and (2). By algebraic manipulation, we can derive the average payoff of cooperators to be
\begin{eqnarray*}
f_{C}=\frac{rc}{N}(N-1)x+\frac{rc}{N}-c-\delta+\frac{a\delta [1-(1-x)^{N}]}{x}.
\end{eqnarray*}
Similarly, the average payoff of defectors is
\begin{eqnarray*}
f_{D}=\frac{rc}{N}(N-1)x-\delta.
\end{eqnarray*}
Accordingly, the difference between the average payoffs of cooperators and defectors can be expressed as $f_{C}-f_{D}=\frac{rc}{N}-c+\frac{a\delta}{x}[1-(1-x)^{N}]$.
Using $1-(1-x)^{N}=x\sum_{k=0}^{N-1}(1-x)^{k}$, we can simplify the above difference in average payoffs as $f_{C}-f_{D}=\frac{rc}{N}-c+a\delta\sum_{k=0}^{N-1}(1-x)^{k}$, which is consistent with previous work \citep{Chen_15_interface}.

\subsection{Feedback-evolving games}

\begin{figure}[t]
\centering
\includegraphics[width=1\linewidth]{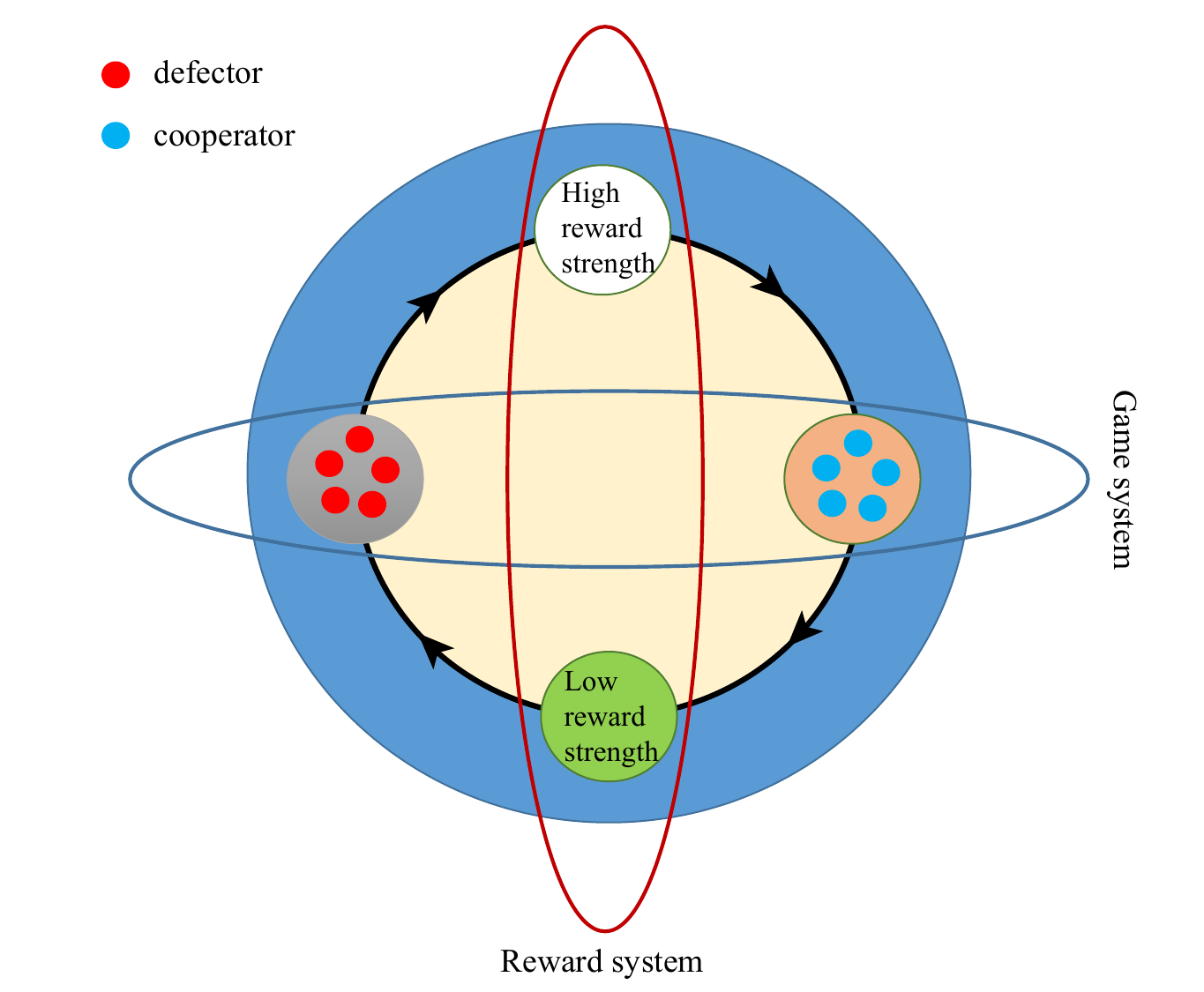}
\caption{Diagram of coupled game-reward system.}
\label{fig1}
\end{figure}

The replicator equations presented above treat reward strength as a critical parameter that influences the evolutionary dynamics of the system. In previous studies, the reward strength was often considered to be a constant value that remains fixed throughout the game \citep{Sasaki_12_PNAS,Chen_15_interface}. However, in real-world scenarios, it is common for the reward strength to vary depending on the number of cooperators in the group. Building upon previous theoretical studies \citep{Weitzpnas2016,Tilman2020}, we construct a feedback-evolving game model in which the strength of reward and the population state form a feedback loop (see Fig.~\ref{fig1}). Specifically, a decrease in the level of cooperation in the population stimulates the institution to increase the reward strength, leading to a rise in the reward strength that promotes the emergence of cooperation. Conversely, an increase in the level of cooperation will prompt the institution to lower the reward strength, which in turn may stimulate individuals to free-ride on the contributions of others.

To describe the impact of strategies on reward strength, we consider a linear feedback form \citep{Weitzpnas2016,Tilman2020}. The specific form of the feedback-evolving game model can be represented as follows 
\begin{equation}\label{equation 6}
\left\{
\begin{aligned}
\epsilon\dot{x} &= x(1-x)[\frac{rc}{N}-c+a\delta\sum_{k=0}^{N-1}(1-x)^{k}],\\
\dot{a} &= (a-\alpha)(\beta-a)[u(1-x)-x],
\end{aligned}
\right.
\end{equation}
where $\epsilon$ represents the relative speed at which individual behavior influences affect the strength of reward. A high value of $\epsilon$ implies that the reward strength can change rapidly in response to changes in the proportion of cooperators or defectors in the group, while a low value of $\epsilon$ indicates that the adjustment of reward strength is relatively slow.  Here, we assume that $0<\epsilon \ll 1$, indicating that the evolution of strategies occurs at a much faster rate compared to the updating of reward strength. Besides, we constrain the strength of rewards to be within the range of $[\alpha, \beta]$, where $\alpha$ represents the minimum reward strength set by the institution, and $\beta$ represents the maximum reward strength. This constraint reflects the practical considerations in implementing incentive mechanisms that balance the need for providing sufficient incentives for cooperation while avoiding overspending. The term $u(1-x)-x$ represents that the reward strength increases at a rate of $u$ as the proportion of defectors in the group increases, decreases with an increase in the proportion of cooperators at a relative rate of one.

To enhance comprehension of the model employed in this study, we present a table summarizing the main parameters and their corresponding definitions. This serves as a useful reference for readers to better grasp the underlying assumptions and mechanisms of the model:
\begin{table}[H]
\centering
\caption{Model parameters and their corresponding definitions}
\begin{tabular}{l|c}
Parameters&Meaning\\\hline
$N$ & Group size \\
$\pi_i$ & Payoff of strategy $i$ in the game \\
$f_i$ & Average payoff of strategy $i$ \\
$r$ & Growth factor of common pool\\
$c$ & Cost of cooperation\\
$\epsilon$ & Relative speed at which individual behavior affects reward strength \\
$\delta$ & Average per capita incentive \\
$x$ & Fraction of cooperation \\
$\alpha$ & Minimum reward strength \\
$\beta$ &  Maximum reward strength\\
$u$ & Rate at which reward strength increases with the proportion of defectors
\\\hline
\end{tabular}\\
\end{table}

In the next section, we will delve into the evolutionary dynamics of the game system described above. Specifically, we will analyze the equilibrium points and their stability to gain insights into the long-term outcomes of the system under different conditions.

\section{Results}


This system (\ref{equation 6}) has at most seven equilibrium points, namely, $(0,\alpha)$, $(0,\beta)$, $(1,\alpha)$, $(1,\beta)$, $(x_{1}, \alpha)$, $(x_{2}, \beta)$, and ($\frac{u}{u+1}$, $a^*$) where $x_{1}$ is solution to the equation $\frac{rc}{N}-c+\alpha\sum_{k=0}^{N-1}(1-x)^{k}=0$, $x_{2}$ is solution to the equation $\frac{rc}{N}-c+\beta\sum_{k=0}^{N-1}(1-x)^{k}=0$, and $a^{*}=\frac{c-rc/N}{\delta\sum_{k=0}^{N-1}(\frac{1}{u+1})^{k}}$. The first four are corner equilibrium points in the phase plane $[0,1]\times [\alpha, \beta]$, the middle two are boundary equilibrium points, and the last one is an interior equilibrium point. In the Appendix, we provide the stability conditions for these equilibrium points based on the sign of the real parts of the eigenvalues of the linearized system's Jacobian matrix \citep{Khalil1996}. Through theoretical analysis, we know that the equilibrium points $(0, \alpha)$ and $(1, \beta)$ are unstable. Based on this, we present five representative evolutionary scenarios.\\

\noindent\textbf{Stable interior equilibrium}\\

When $\delta>\frac{c-rc/N}{\sum_{k=0}^{N-1}(\frac{1}{u+1})^{k}}$, we know that system (\ref{equation 6}) has an interior equilibrium point. Through theoretical analysis, we are able to determine that it is stable (Detailed theoretical analysis is provided in the Appendix). In addition, when $\frac{c-rc/N}{N\alpha}<\delta<\frac{c-rc/N}{\alpha}$, the boundary equilibrium point $(x_{1}, \alpha)$ exists and it is unstable when $u-ux_{1}-x_{1}>0$.
Furthermore, considering that $\delta>c-\frac{rc/N}{N\alpha}>c-\frac{rc/N}{N\beta}$, we know that the corner equilibrium point $(0, \beta)$ is unstable. At the same time, due to $\delta<\frac{c-rc/N}{\alpha}$, we can determine that the corner equilibrium point $(1,\alpha)$ is also unstable. Therefore, all four corner equilibrium points are unstable.

\begin{figure}[t]
\centering
\includegraphics[width=1\linewidth]{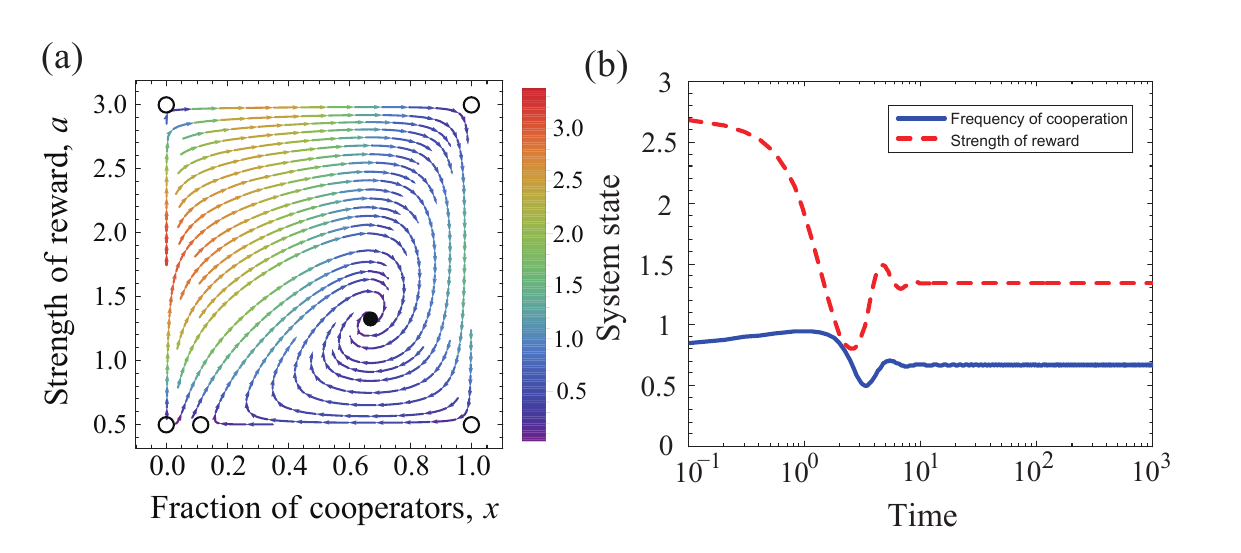}
\caption{Stable interior equilibrium point.
The replicator equation's phase portraits with diverse initial conditions are illustrated in Panel (a). Stable equilibrium points are represented by solid dots, whereas unstable equilibrium points are depicted as empty circles. Evolutionary direction is indicated by arrows with color used to represent the magnitude of the gradient of selection. Panel (b) displays the temporal evolution of strategy frequency and punishment intensity with a specified initial value. The red dashed line represents the strength of reward, while the blue solid line represents the frequency of cooperation. Parameter values are $N=5, F=3, c=1, \epsilon=0.1, \delta=0.2, \alpha=0.5, \beta=3$, and $u=2$.}
\label{fig2}
\end{figure}

Figure~\ref{fig2} provides numerical examples to verify the theoretical analysis mentioned above. Specifically, Fig.~\ref{fig2}(a) shows the phase portrait of the replicator system when the parameters satisfy the aforementioned conditions. Unstable equilibrium points are represented by empty circles, stable equilibrium points by solid dots, while arrows indicate the direction of evolution and the color of arrows depict the magnitude of gradient of selection, red indicating the highest gradient value and blue representing the lowest gradient value. We can observe that there are six equilibrium points in the system, with the interior equilibrium point being stable while the other five equilibrium points are unstable. All interior trajectories ultimately converge to this stable equilibrium point. In Fig.~\ref{fig2}(b), we present the temporal evolution of the cooperation frequency and reward strength with given initial conditions. We observe that the frequency of cooperation and reward strength gradually decrease at first, then increase, and finally decline to stabilize at a certain fixed value.\\

\noindent\textbf{Full-cooperation state with low-intensity reward}\\

\begin{figure}[t]
\centering
\includegraphics[width=1\linewidth]{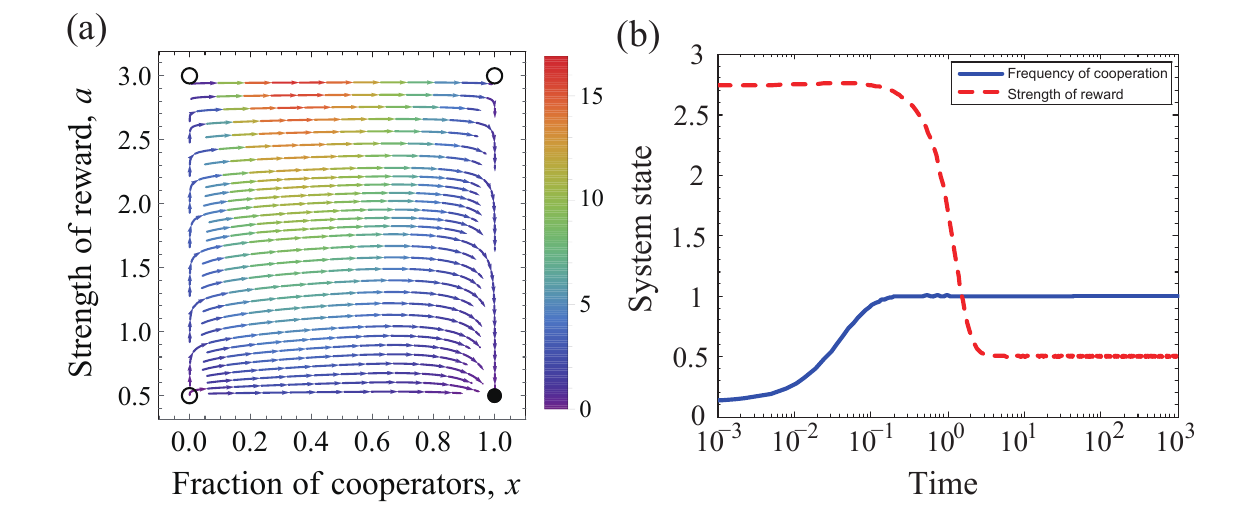}
\caption{Full-cooperation state with low-intensity reward. Panel (a) exhibits phase portraits of the replicator equation. Panel (b) displays the temporal evolution of cooperation frequency and reward strength for a specific initial condition. Parameter values are $N=5, F=3, c=1, \epsilon=0.1, \delta=1, \alpha=0.5, \beta=3$, and $u=2$.}
\label{fig3}
\end{figure}

When $\delta<\frac{c-rc/N}{\sum_{k=0}^{N-1}(\frac{1}{u+1})^{k}}$, it can be analyzed that the system does not have interior equilibrium point. On the other hand, when $\delta>\frac{c-rc/N}{\alpha}$, the corner equilibrium point $(1, \alpha)$ is stable. In this case, both boundary equilibrium points $(x_{1}, \alpha)$ and $(x_{2}, \beta)$ do not exist. In summary, the system (\ref{equation 6}) has four corner equilibrium points, namely $(0, \alpha)$, $(0, \beta)$, $(1, \alpha)$, and $(1, \beta)$. Through analysis, it can be concluded that only the corner equilibrium point $(1, \alpha)$ is stable, while the other three corner equilibrium points are unstable, which means that maintaining a state of full cooperation state only requires providing the minimum level of reward strength.

We present numerical examples in Fig.~\ref{fig3} to validate the theoretical analysis mentioned above. From Fig.~\ref{fig3}(a), we can observe that there are four corner equilibrium points in the phase plane, among which (1, 0.5) is a stable equilibrium point, and the other three are unstable. Trajectories starting from different initial points converge to (1, 0.5) eventually. In Fig.~\ref{fig3}(b), we show the time evolution of system states under specific initial conditions. We can observe that the reward strength initially remains at a high level, then gradually decreases, and finally stabilizes at the minimum level ($a=0.5$). Meanwhile, the frequency of cooperation gradually increases and eventually stabilizes at 1.\\
\begin{figure}[t]
\centering
\includegraphics[width=1\linewidth]{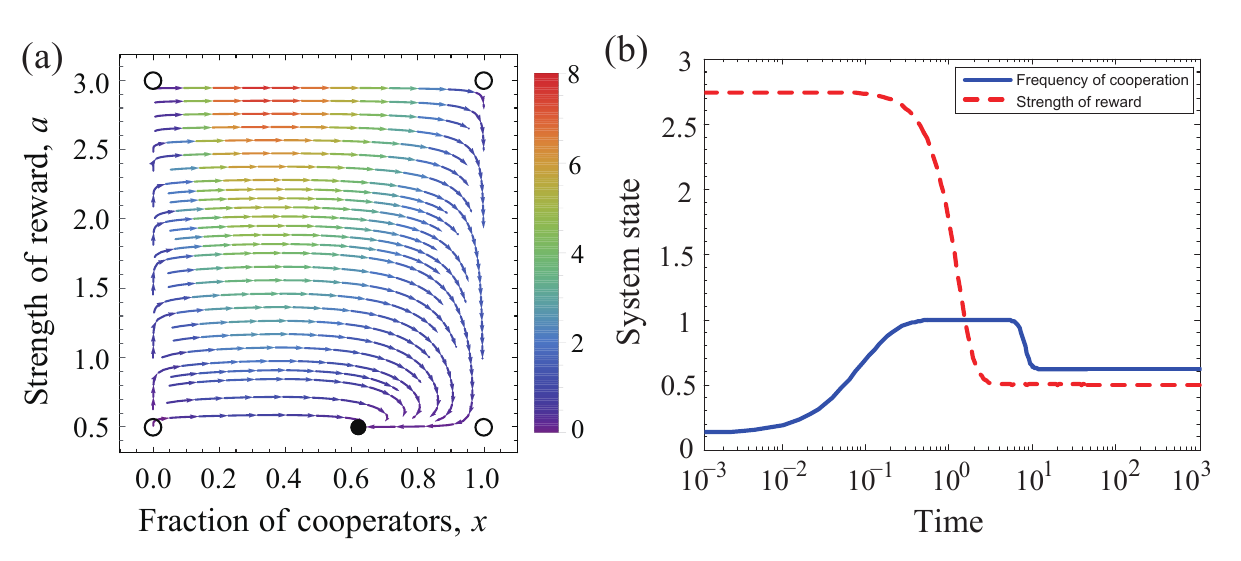}
\caption{Stable boundary equilibrium point $(x_{1}, \alpha)$. Panel (a) presents the phase plane diagram of the replicator system. Panel (b) shows the time evolution of the system state. Parameter values are $N=5, F=3, c=1, \epsilon=0.1, \delta=0.5, \alpha=0.5, \beta=3$, and $u=0.6$.}
\label{fig4}
\end{figure}

\noindent\textbf{Moderate level of cooperation with low-intensity reward}\\

When $\delta<\frac{c-rc/N}{\sum_{k=0}^{N-1}(\frac{1}{u+1})^{k}}$, the interior equilibrium point does not exist. When $\frac{c-rc/N}{N\alpha}<\delta<\frac{c-rc/N}{\alpha}$, the boundary equilibrium point $(x_{1}, \alpha)$ exists, and we can determine that it is stable when $u - ux_{1} - x_{1} < 0$. The other boundary equilibrium point $(x_{2}, \beta)$ does not exist when $\delta > \frac{c - rc/N}{\beta}$. Therefore, the system has five equilibrium points, including four corner equilibrium points and one boundary equilibrium point. Considering $\delta < \frac{c - rc/N}{\alpha}$, we can know that the equilibrium point $(1, \alpha)$ is unstable. At the same time, since $\delta > \frac{c - rc/N}{N\alpha}>\frac{c - rc/N}{N\beta}$, we know that the equilibrium point $(0, \beta)$ is also unstable. Therefore, when $u - ux_{1} - x_{1} < 0$, only the boundary equilibrium point $(x_{1}, \alpha)$ is stable, and the other equilibrium points are unstable. This means that a moderate level of cooperation can be sustained by providing the minimum level of reward strength.

In Fig.~\ref{fig4}, we present a specific numerical example to validate the theoretical analysis results mentioned above. As shown in Fig.~\ref{fig4}(a), there are five equilibrium points in the phase plane, including four corner equilibrium points and one boundary equilibrium point. We can observe that trajectories starting from different initial points converge to the boundary equilibrium point $(x_{1}, \alpha)$. Fig.~\ref{fig4}(b) shows the time evolution of the system state. We can observe that the reward strength initially remains at a high level, then gradually decreases, and finally stabilizes at the minimum level of reward strength, which is 0.5. Meanwhile, the frequency of cooperation initially increases, then maintains a sufficient level of cooperation for some time before gradually decreasing and finally stabilizing near 0.6.\\

\noindent\textbf{Moderate level of cooperation with high-intensity reward}\\

\begin{figure}[t]
\centering
\includegraphics[width=1\linewidth]{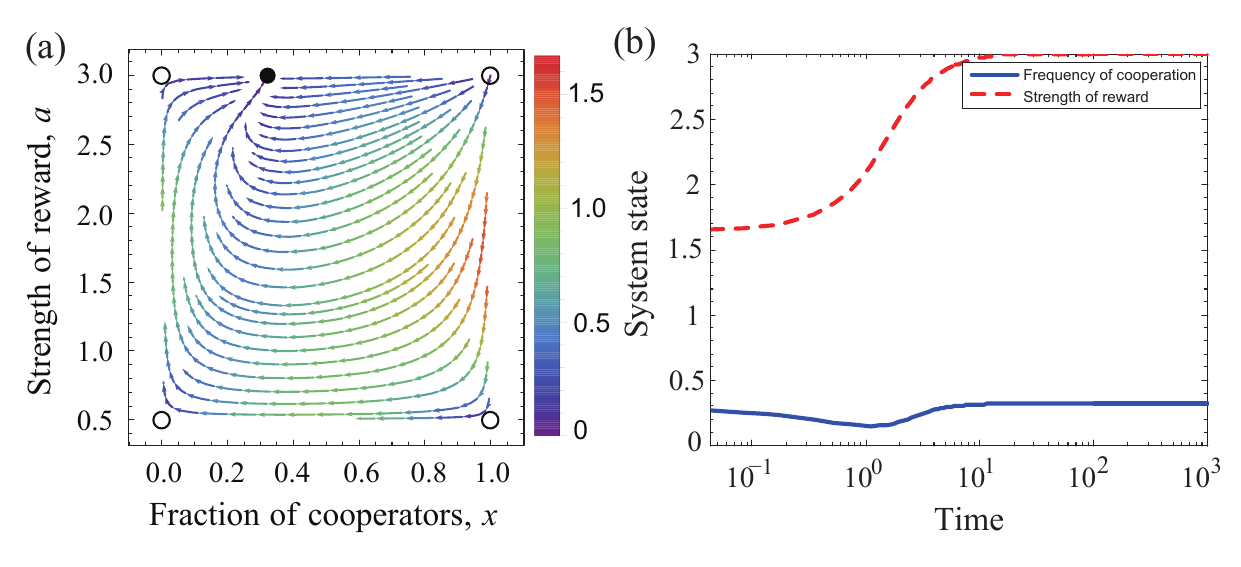}
\caption{Stable boundary equilibrium point $(x_{2}, \beta)$. Panel (a) presents the phase plane diagram of the replicator system. Panel (b) shows the time evolution of the system state. Parameter values are $N=5, F=3, c=1, \epsilon=0.1, \delta=0.05, \alpha=0.5, \beta=3$, and $u=0.6$.}
\label{fig5}
\end{figure}

When $\frac{c-rc/N}{N\beta}<\delta<\frac{c-rc/N}{\beta}$, the boundary equilibrium point $(x_{2}, \beta)$ exists, and we know that it is stable when $u - ux_{2} - x_{2} < 0$. In addition, when $\delta < \min\{\frac{c-rc/N}{\sum_{k=0}^{N-1}(\frac{1}{u+1})^{k}}, \frac{c-rc/N}{N\alpha}\}$, we know that both the interior equilibrium point $(\frac{u}{1+u}, \frac{c-rc/N}{\delta\sum_{k=0}^{N-1}(\frac{1}{u+1})^{k}})$ and the boundary equilibrium point $(x_{1}, \alpha)$ do not exist. Thus the system has five fixed point. Considering $\delta < \frac{c - rc/N}{\beta}<\frac{c - rc/N}{\alpha}$, we can know that the equilibrium point $(1, \alpha)$ is unstable. At the same time, since $\delta > \frac{c - rc/N}{N\beta}$, we know that the equilibrium point $(0, \beta)$ is also unstable. Therefore, when $u - ux_{2} - x_{2} < 0$, only the boundary equilibrium point $(x_{2}, \beta)$ is stable, and the other equilibrium points are unstable, which means that a moderate level of cooperation can be sustained, but it requires providing a sufficiently high level of reward strength.

Fig.~\ref{fig5}(a) presents the phase diagram of replicator system (\ref{equation 6}). We can observe that there are five equilibrium points in the phase plane, including four corner equilibrium points and one boundary equilibrium point $(x_{2}, \beta)$. Only the boundary equilibrium point is stable, and all trajectories eventually converge to this stable equilibrium point. In Fig.~\ref{fig5}(b), we present a specific example. We can observe that the reward strength gradually increases over time, reaching its maximum value of 3. Meanwhile, the cooperation frequency initially decreases and then increases, finally stabilizing near 0.3.\\

\noindent\textbf{Full defection with high-intensity reward}\\

\begin{figure}[t]
\centering
\includegraphics[width=1\linewidth]{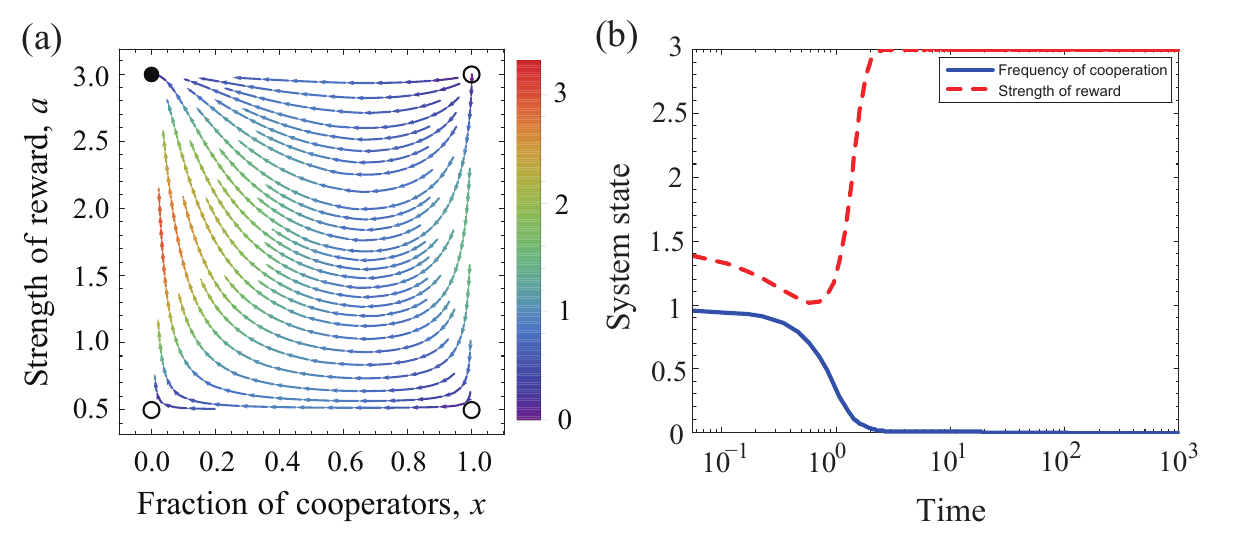}
\caption{Full defection with high-intensity reward. Panel (a) presents the phase plane diagram of the replicator system. Panel (b) shows the time evolution of the system state. Parameter values are $N=5, F=3, c=1, \epsilon=0.1, \delta=1/75, \alpha=0.5, \beta=3$, and $u=2$.}
\label{fig6}
\end{figure}

When $\delta < \min\{\frac{c-rc/N}{\sum_{k=0}^{N-1}(\frac{1}{u+1})^{k}}, \frac{c-rc/N}{N\beta}\}$, we know that both the interior equilibrium point and two boundary equilibrium point do not exist. In this case, the system has four corner equilibrium points.
We can easily obtain that the equilibrium point $(1, \alpha)$ is unstable because $\delta < \frac{c-rc/N}{N\beta}<\frac{c - rc/N}{\alpha}$. While the equilibrium point $(0, \beta)$ is stable because $\delta < \frac{c-rc/N}{N\beta}$. This means that even with a very high level of reward strength, it is still not possible to promote the emergence of cooperation.

In Fig.~\ref{fig6}, we provide a numerical example when the parameters satisfy the above conditions. We can observe that there are four corner equilibrium points in the phase plane, among which only (0,3) is stable. All interior trajectories eventually converge to (0,3) regardless of the initial conditions (see Fig.~\ref{fig6}(a)). In Fig.~\ref{fig6}(b), we present the time evolution of the system state when the initial conditions are fixed. We can observe that the reward strength initially decreases, then gradually increases, and finally stabilizes at the upper limit of reward strength. Meanwhile, the frequency of cooperation gradually decreases and eventually stabilizes at 0.

\begin{figure}[t]
\centering
\includegraphics[width=1\linewidth]{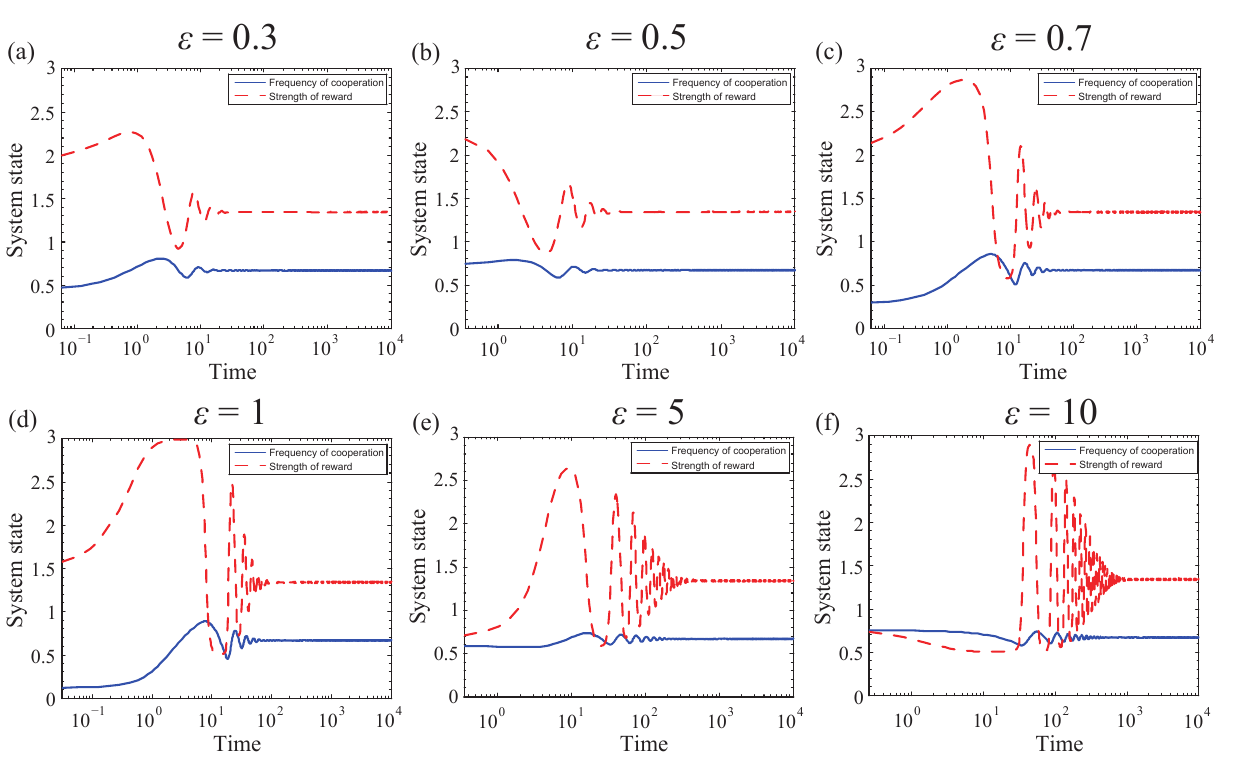}
\caption{Invariance of system dynamics given change in the feedback speed. The parameter $\epsilon$ is being varied from 0.3 to 10 given cases where the dynamics are expected to lead to a stable interior fixed point. Parameter values are $N=5, F=3, c=1, \delta=0.2, \alpha=0.5, \beta=3$, $u=2$, $\epsilon=0.3$ in panel (a); $\epsilon=0.5$ in panel (b); $\epsilon=0.7$ in panel (c); $\epsilon=1$ in panel (d); $\epsilon=5$ in panel (e); $\epsilon=10$ in panel (f).}
\label{fig7}
\end{figure}

Finally, we are interested in how feedback speed affects the coevolutionary outcomes when a stable interior equilibrium point is reached in the system. As shown in Fig.~\ref{fig7}, when the feedback speed $\epsilon$ changes from 0.3 to 10, a stable interior equilibrium point still exists in the system. This implies that when the feedback speed varies within a certain range, the qualitative outcome remains independent of the feedback speed.

\section{Conclusions}

Reward is an effective way to promote the emergence of cooperation. By providing incentives to individuals or groups who exhibit cooperative behavior, rewards can facilitate and reinforce collaborative efforts and help to improve the welfare of the entire society. Previous studies have investigated the role of different types of rewards in promoting cooperation, with particular attention given to institutional rewards in recent years \citep{sun2021iscience,Sasaki_12_PNAS}. However, studies involving institutional rewards typically assume that the intensity of the reward is constant and does not vary with changes in the game environment or state. In this work, we have proposed an adaptive institutional reward strategy and established a coevolutionary dynamics model, which extends previous research by incorporating the dynamic nature of reward systems, allowing for a more realistic representation of how reward structures can evolve over time in response to changes in population behavior. We have revealed five representative evolutionary outcomes, including stable interior equilibrium point, the minimum reward intensity for full cooperation, the minimum reward intensity for moderate levels of cooperation, the maximum reward intensity for moderate levels of cooperation, and the maximum reward intensity for full defection.

We have demonstrated that the replicator system can produce a stable interior equilibrium point, implying that a constant reward value can maintain cooperation at a certain level. In the dynamic feedback design between reward intensity and frequency of cooperation, the interior stable point is an important outcome. As the reward intensity and level of cooperation gradually approach equilibrium, the system tends to stabilize and form an internal stable point. This interior stable point may be crucial for achieving long-term cooperation since it can help prevent the system from deviating or collapsing. We have verified that the feedback speed does not alter this evolutionary outcome.

We have also revealed that the dynamic system coupling reward intensity and population state can achieve an ideal state where the population can maintain full cooperation while the reward intensity is kept at a minimum value. Previous theoretical research has examined the effect of institutional rewards on the evolution of cooperation under the condition of constant reward intensity based on public goods games \citep{Chen_15_interface}. Theoretical results indicate that when the incentive is sufficiently large, i.e., $\delta>\frac{c-rc/N}{a}$, the state of full cooperation can be achieved. By releasing the assumption of a constant reward intensity, and introducing an adaptive reward strategy, we have revealed that the achievement of the state of full cooperation with a minimum reward intensity requires $\delta>\frac{c-rc/N}{\alpha}$. It is evident that it depends on the value of the minimum reward intensity.

It has been suggested that an adaptive adjustment of reward strength based on the proportion of cooperators can be considered a manifestation of the principle of diminishing marginal returns, which stipulates that the effectiveness of a given reward diminishes as more individuals receive it. Thus, when there are many cooperators, reducing the reward strength would allow for a more efficient allocation of resources. Conversely, when the proportion of cooperators is low, increasing the reward strength can serve to incentivize initial participation and promote cooperation. By adapting the strength of rewards based on the proportion of cooperators, we can achieve a balance between providing sufficient incentives for cooperation and avoiding excessive expenditures. These insights shed light on how the adaptive adjustment of reward strength influences the evolution of cooperation in various social contexts.

To sum up, our investigation has shed light on the crucial role of the feedback loop between reward intensity and population state in shaping cooperative behavior. Further investigations into various forms of feedback and group structures hold great potential for advancing the development of control strategies that are both efficient and effective in complex systems. Specifically, future research can focus on examining approaches to attain system stability and controllability through exponential feedback mechanisms \citep{Liu_22_csf,Liu_23_elife}, while considering diverse population structures \citep{Perc_JRSInterface_13,xu2023scis} and the optimal institutional incentives \citep{wang2019cnsns,Cimpeanu_DGA_2023}. By delving into these areas, we can enhance our understanding of cooperative dynamics and pave the way for more refined and adaptable control strategies in complex systems.
 \\

\noindent\textbf{\large Appendix}\\

\noindent\textbf{Equilibrium analysis of the coupled system}\\
The coupled system that describes the dynamics of the feedback-evolving game is given as follows:
\begin{equation*}
\left\{
\begin{aligned}
\epsilon\dot{x} &= x(1-x)[\frac{rc}{N}-c+a\delta\sum_{k=0}^{N-1}(1-x)^{k}],\\
\dot{a} &= (a-\alpha)(\beta-a)[u(1-x)-x].
\end{aligned}
\right.
\end{equation*}
This dynamical system has a maximum of seven fixed points, which include all four corners of the domain $[0,1]\times [\alpha,\beta]$, the boundary fixed point $(x_{1}, \alpha)$ and $(x_{2}, \beta)$, and the interior fixed point ($\frac{u}{u+1}$, $a^*$) where $x_{1}$ satisfies the equation $\frac{rc}{N}-c+\alpha\delta\sum_{k=0}^{N-1}(1-x)^{k}=0$, $x_{2}$ satisfies the equation $\frac{rc}{N}-c+\beta\delta\sum_{k=0}^{N-1}(1-x)^{k}=0$, and $a^{*}=\frac{c-rc/N}{\delta\sum_{k=0}^{N-1}(\frac{1}{u+1})^{k}}$.

The system always has four corner equilibrium points. However, there are also two boundary equilibrium points that occur at the boundaries of the domain and may exist under certain conditions. 
In the following, we analyze the conditions for the existence of two boundary equilibrium points $(x_{1}, \alpha)$ and $(x_{2}, \beta)$. Here we set that $G_{1}(x)=\frac{rc}{N}-c+\alpha\delta\sum_{k=0}^{N-1}(1-x)^{k}$ and $G_{2}(x)=\frac{rc}{N}-c+\beta\delta\sum_{k=0}^{N-1}(1-x)^{k}$, then $G_{1}'(x)=-\alpha\delta\sum_{k=1}^{N-1}k(1-x)^{k-1}$ and $G_{2}'(x)=-\beta\delta\sum_{k=1}^{N-1}k(1-x)^{k-1}$, which are negative for $x\in(0,1)$. 
Considering that $G_{1}(0)=\frac{rc}{N}-c+N\alpha\delta$ and $G_{1}(1)=\frac{rc}{N}-c+\delta\alpha$, we can derive the conditions for the existence of the boundary equilibrium point $(x_{1}, \alpha)$ as $\frac{c-rc/N}{N\alpha}<\delta<\frac{c-rc/N}{\alpha}$. Similarly, we can derive the conditions for the existence of the boundary equilibrium point $(x_{2}, \beta)$ as $\frac{c-rc/N}{N\beta}<\delta<\frac{c-rc/N}{\beta}$. Therefore, the condition for the existence of both boundary equilibrium points $(x_{1}, \alpha)$ and $(x_{2}, \beta)$ is $\frac{c-rc/N}{N\alpha}<\delta<\frac{c-rc/N}{\beta}$.

Besides, the condition for the existence of the interior equilibrium point is $0<\frac{c-rc/N}{\delta\sum_{k=0}^{N-1}(\frac{1}{1+u})^{k}}<1$. 

For convenience, we let $h(x,a)=\frac{1}{\epsilon}x(1-x)[\frac{rc}{N}-c+a\delta\sum_{k=0}^{N-1}(1-x)^{k}]$ and $g(x,a)=(a-\alpha)(\beta-a)[u(1-x)-x]$. In order to determine the stability of equilibrium points in a system, we can investigate the Jacobian matrix $J$ \citep{Khalil1996}. The elements of this matrix are given by the following formula:
\begin{eqnarray}
J =
\begin{pmatrix}
\frac{\partial h}{\partial x} & \frac{\partial h}{\partial a} \\
\frac{\partial g}{\partial x} & \frac{\partial g}{\partial a}
\end{pmatrix},
\end{eqnarray}
where
\begin{eqnarray}
\left\{
\begin{aligned}
\frac{\partial h}{\partial x} &= \frac{1}{\epsilon}(1-2x)(\frac{rc}{N}-c)+\frac{1}{\epsilon}a\delta[(N+1)(1-x)^{N}-1],\\
\frac{\partial h}{\partial a} &= \frac{1}{\epsilon}(1-x)\delta[1-(1-x)^{N}],\\
\frac{\partial g}{\partial x} &=(a-\alpha)(\beta-a)(-u-1),\\
\frac{\partial g}{\partial a} &=(\beta-2a+\alpha)[u(1-x)-x].
\end{aligned}
\right.
\end{eqnarray}

We will now proceed with an analysis of the stability of each equilibrium point. 

For $(x,a) = (0,\alpha)$, the Jacobian is 
\begin{align}
J(0,\alpha) =
\begin{pmatrix}
\frac{1}{\epsilon}(\frac{rc}{N}-c)+\frac{1}{\epsilon}N\alpha\delta & 0 \\
0 & (\beta-\alpha)u
\end{pmatrix},
\end{align}
we can know that it is unstable since $(\beta-\alpha)u>0$.

For $(x,a) = (0,\beta)$, the Jacobian is 
\begin{align}
J(0,\beta) =
\begin{pmatrix}
\frac{1}{\epsilon}(\frac{rc}{N}-c)+\frac{1}{\epsilon}N\beta\delta & 0 \\
0 & (\alpha-\beta)u
\end{pmatrix},
\end{align}
we can know that it is stable when $\frac{1}{\epsilon}(\frac{rc}{N}-c)+\frac{1}{\epsilon}N\beta\delta< 0$.

For $(x,a) = (1,\alpha)$, the Jacobian is 
\begin{align}
J(1,\alpha) =
\begin{pmatrix}
-\frac{1}{\epsilon}(\frac{rc}{N}-c)-\frac{1}{\epsilon}\alpha\delta & 0 \\
0 & -(\beta-\alpha)
\end{pmatrix},
\end{align}
we can know that it is stable when $-\frac{1}{\epsilon}(\frac{rc}{N}-c)-\frac{1}{\epsilon}\alpha\delta<0$.

For $(x,a) = (1,\beta)$, the Jacobian is 
\begin{align}
J(1,\beta) =
\begin{pmatrix}
-\frac{1}{\epsilon}(\frac{rc}{N}-c)-\frac{1}{\epsilon}\beta\delta & 0 \\
0 & -(\alpha-\beta)
\end{pmatrix},
\end{align}
we can know that it is unstable since $-(\alpha-\beta)>0$.

For $(x,a) = (x_{1},\alpha)$, the Jacobian is 
\begin{align}
J(x_{1},\alpha) =
\begin{pmatrix}
a_{11} & \frac{1}{\epsilon}(1-x_{1})\delta[1-(1-x_{1})^{N}] \\
0 & (\beta-\alpha)(u-ux_{1}-x_{1})
\end{pmatrix},
\end{align}
where $a_{11}=\frac{1}{\epsilon}x_{1}(1-x_{1})[-\alpha\delta\sum_{k=0}^{N-1}k(1-x_{1})^{k-1}]<0$ and this equilibrium point is stable when $u-ux_{1}-x_{1}<0$.

For $(x,a) = (x_{2},\beta)$, the Jacobian is 
\begin{align}
J(x_{2},\beta) =
\begin{pmatrix}
\bar{a}_{11} & \frac{1}{\epsilon}(1-x_{2})\delta[1-(1-x_{2})^{N}] \\
0 & (\alpha-\beta)(u-ux_{2}-x_{2})
\end{pmatrix},
\end{align}
where $\bar{a}_{11}=\frac{1}{\epsilon}x_{2}(1-x_{2})[-\alpha\delta\sum_{k=0}^{N-1}k(1-x_{2})^{k-1}]<0$, and we can know that it is stable when $u-ux_{2}-x_{2}<0$.

For $(x,a) = (\frac{u}{u+1}, a^*)$, the Jacobian is 
\begin{align}
J(\frac{u}{u+1}, a^*) =
\begin{pmatrix}
a_{11}^{*} & \frac{1}{\epsilon}\frac{1}{u+1}\delta[1-(\frac{1}{u+1})^{N}] \\
(a^{*}-\alpha)(\beta-a^{*})(-u-1) & 0
\end{pmatrix},
\end{align}
where $a_{11}^{*}=\frac{1}{\epsilon}\frac{u}{u+1}\frac{1}{u+1}[-\alpha\delta\sum_{k=0}^{N-1}k(\frac{1}{u+1})^{k-1}]<0$. We can get tr $(J)=a_{11}^{*}<0$ and det $(J)=\frac{1}{\epsilon}\delta(a^{*}-\alpha)(\beta-a^{*})[1-(\frac{1}{u+1})^N]>0$, thus this point is stable.

\section*{CRediT authorship contribution statement}
\noindent Shijia Hua: Investigation, Discussing, Writing - review \& editing.\\
Linjie Liu: Conceptualization, Discussing, Writing - review \& editing.

\section*{Declaration of competing interest}
The authors declare that they have no known competing financial interests or personal relationships that could have appeared
to influence the work reported in this paper.

\section*{Acknowledgments}
This research was supported by the Natural Science Foundation of Shaanxi (Grant No. 2023-JC-QN-0791) and the Fundamental Research Funds of the Central Universities of China (Grants No. 2452022012, 2452022144).\\



\end{document}